\documentclass[12pt,preprint]{aastex}

\newcommand{\bvec}[1]{\textbf{#1}}

\shorttitle{Weak-lensing analysis of Cl1226}
\shortauthors{Jee \& Tyson}

\begin{document}

\title{DARK MATTER IN THE GALAXY CLUSTER CL J1226+3332 AT $Z=0.89$} 

\author{M. JAMES JEE\altaffilmark{1} and J. ANTHONY TYSON\altaffilmark{1} }

\begin{abstract}
We present a weak-lensing analysis of the galaxy cluster CL J1226+3332 at $z\simeq0.89$
using {\it Hubble Space Telescope} Advanced Camera for Surveys images. 
The cluster is the hottest ($>10~$keV), most X-ray luminous system at $z>0.6$ known to date.
The relaxed X-ray morphology, as well as its high temperature, is unusual at such a high redshift.
Our mass reconstruction shows that on a large scale the dark matter distribution
is consistent with a relaxed system with no significant substructures. However, on a small scale the cluster core
is resolved into two mass clumps highly correlated with the cluster galaxy distribution.
The dominant mass clump lies close to the brightest cluster galaxy whereas the other less massive clump is located $\sim40\arcsec$ ($\sim310$ kpc)
to the southwest.
Although this secondary mass clump does not show an excess in the X-ray surface brightness, the gas temperature
of the region is much higher ($12\sim18$ keV) than those of the rest. We propose a scenario in which
the less massive system has already passed through the main cluster and the X-ray gas has been stripped during this passage.
The elongation of the X-ray peak toward the southwestern mass clump is also supportive of this possibility.
We measure significant tangential shears out to the field boundary ($\sim1.5$ Mpc), which
are well described by an Navarro-Frenk-White profile with a concentration parameter of $c_{200}=2.7\pm0.3$ and
a scale length of $r_s=78\arcsec\pm19\arcsec$ ($\sim600$ kpc) with $\chi^2/$d.o.f=1.11.
Within the spherical volume $r_{200}=1.6$~Mpc, the total mass of the cluster becomes
$M(r<r_{200})=(1.4\pm0.2)\times10^{15}~M_{\sun}$. 
Our weak-lensing analysis confirms that CL1226+3332 is indeed the most massive cluster known to date at $z>0.6$.
\end{abstract}

\altaffiltext{1}{Department of Physics, University of California, Davis, One Shields Avenue, Davis, CA 95616}


\section{INTRODUCTION \label{section_introduction}}
In the hierarchical structure formation paradigm with cold dark matter (CDM),
galaxy clusters 
grow through multiple mergers between groups and smaller clusters of galaxies.
Accordingly, on average galaxy clusters found at higher redshifts should be less massive and
more irregular. Despite many unresolved issues in detail on how these structures grow over time,
because cluster assembly on a large scale is governed by
CDM only subject to gravity, it is possible to quantitatively predict a cluster
mass function for a given cosmology at a specific epoch through either analytic approaches 
or numerical simulations.
Therefore, comparison of cluster mass functions today and at high redshift
has been a powerful tool in constraining cosmological parameters.

Although large samples of high-redshift clusters have been compiled in the past decade
through extensive surveys, the principal ambiguity yet to be
resolved is how to relate the measured cluster properties to the mass of the system 
in a quantitatively robust way. This issue becomes more important for the most massive clusters
at high redshifts ($z\sim1$). The statistics of these rare systems are extremely sensitive
to the matter content ($\Omega_M$) of the universe and its fluctuation ($\sigma_8$).
A historic example is MS1054-0321 (hereafter MS1054), the hottest cluster in the Einstein 
Medium Sensitivity Survey (EMSS) at $z=0.83$. The mere existence of such a massive cluster
is thought to be problematic in the $\Omega_M=1$ universe and has been frequently used as an
argument for low $\Omega_M<1$ values (Bahcall \& Fan 1998; Donahue et al. 1998; Jeltema et al. 2001). It is interesting to note, however, that
the intracluster medium (ICM) temperature of MS1054 has been debated 
(Donahue et al. 1998; Jeltema et al. 2001; Joy et al. 2001; Vikhlinin et al. 2002; Tozzi et al. 2003; Gioia et al. 2004; Jee et al. 2005b; Branchesi et al. 2007)
and this challenges
the mass estimate based on X-ray observations alone. Moreover, the complicated substructures
of MS1054 ubiquitously present in optical, X-ray, and weak-lensing observations make
the validity of the hydrostatic equilibrium assumption (commonly questioned even for a relaxed
system) more open to doubts. However, weak-lensing analyses, which
do not depend on the dynamical state of the cluster, from {\it Hubble Space Telescope (HST)} Wide Field Planetary Camera 2 (WFPC2) (Hoekstra et al. 2000)
and Advanced Camera for Surveys (ACS) (Jee et al. 2005b) show that MS1054 is indeed a massive system at $z=0.83$.

Presumably, the galaxy cluster CL J1226.9+3332 (hereafter CL1226) at $z=0.89$ is much more massive than MS1054.
The recent measurements of the X-ray temperature $10.4\pm0.6$ keV
and the total X-ray luminosity ($L_X = 5.12\pm 0.12~\mbox{erg}~\mbox{s}^{-1}$) of the cluster by Maughan et al. (2007)
show that the cluster is the hottest, most luminous system at $z>0.6$ known to date. These X-ray
properties suggest that CL1226 might be perhaps the most massive $z>0.6$ structure as well; the detailed
3-dimensional mass structure analysis of Maughan et al. (2007) derives an enclosed mass 
$M(r<0.88 \mbox{Mpc})=5.2_{-0.8}^{+1.0}\times 10^{14}~M_{\sun}$.
The relaxed X-ray morphology of the cluster is also remarkable; the cluster's X-ray surface brightness distribution is symmetric
with a single X-ray peak in spatial agreement with the brightest cluster galaxies. This
is in stark contrast with the X-ray images of CL0152-0152 (hereafter CL0152; Maughan et al. 2003) and MS1054 (Jeltema et al. 2001). 
These two high-redshift clusters at a similar
redshift of $z\sim0.8$ show multiple peaks in their X-ray images indicative of their active stage of formation. 
Within the hierarchical structure formation paradigm, a virialized structure like CL1226 with such a high mass
is extremely rare at $z=0.89$, when the age of the universe is less than half its current value.

In this paper, we present a weak-lensing study of CL1226 with HST/ACS
images. By analyzing weak distortions of background galaxies behind the cluster, we aim to address
the following issues. First, we will examine how the lensing mass compares the X-ray value. Although the analysis
of Maughan et al. (2007) using both $Chandra$ and XMM-$Newton$ data was careful and certainly more sophisticated 
than the common isothermal $\beta$ model approach, the X-ray method is by nature still dependent on the validity
of the hydrostatic equilibrium assumption. Because the relaxed X-ray morphology of the system supports this
hypothesis, it provides an interesting opportunity to compare weak-lensing and X-ray estimates on a fair basis.
In addition, Maughan et al. (2007) found that their X-ray mass is $\sim30$\% lower than the predicted $M-T$ (Vikhlinin et al. 2006) and
$Y_X-T$ (Kravtsov et al. 2006) relation. If the cause of this departure is an on-going merger activity as suggested
by Maughan et al. (2007), our weak-lensing mass estimate should improve the cluster's scaling relation.
Second, we will investigate if our two-dimensional mass reconstruction uncovers any substructure not observed
in X-rays. In Jee et al. (2005a; 2005b), we witnessed that the weak-lensing mass in CL0152 and MS1054 traces the cluster galaxy distribution
very closely whereas the X-ray maps often do not always show the same details. Therefore, it is probable that
the weak-lensing analysis of CL1226 might reveal some significant substructures that have not been detected by X-rays.
In particular, Maughan et al. (2007) noted that the ICM temperature of the region $\sim40\arcsec$ to the
southwest of the X-ray peak is much higher ($12-18~$keV) than the average value ($\sim10.4$ keV). Although
this hot region is correlated with the cluster galaxy distribution, the structure does not stand out
in X-ray surface brightness. Perhaps either the mass associated with this galaxy group is not significant
enough to produce X-ray overdensity, or the X-ray gas has been stripped in a previous pass-through. 
Our weak-lensing measurement of the mass of this substructure will allow us to address this question.

Throughout the paper, we use a $(h,\Omega_M,\Omega_{\Lambda})=(0.7,0.3,0.7)$ cosmology. All the quoted uncertainties
are at the 1-$\sigma$ (68\%) level.

\section{OBSERVATIONS \label{section_obs}}

\subsection{Data Reduction and Object Detection}
\begin{figure}
\plotone{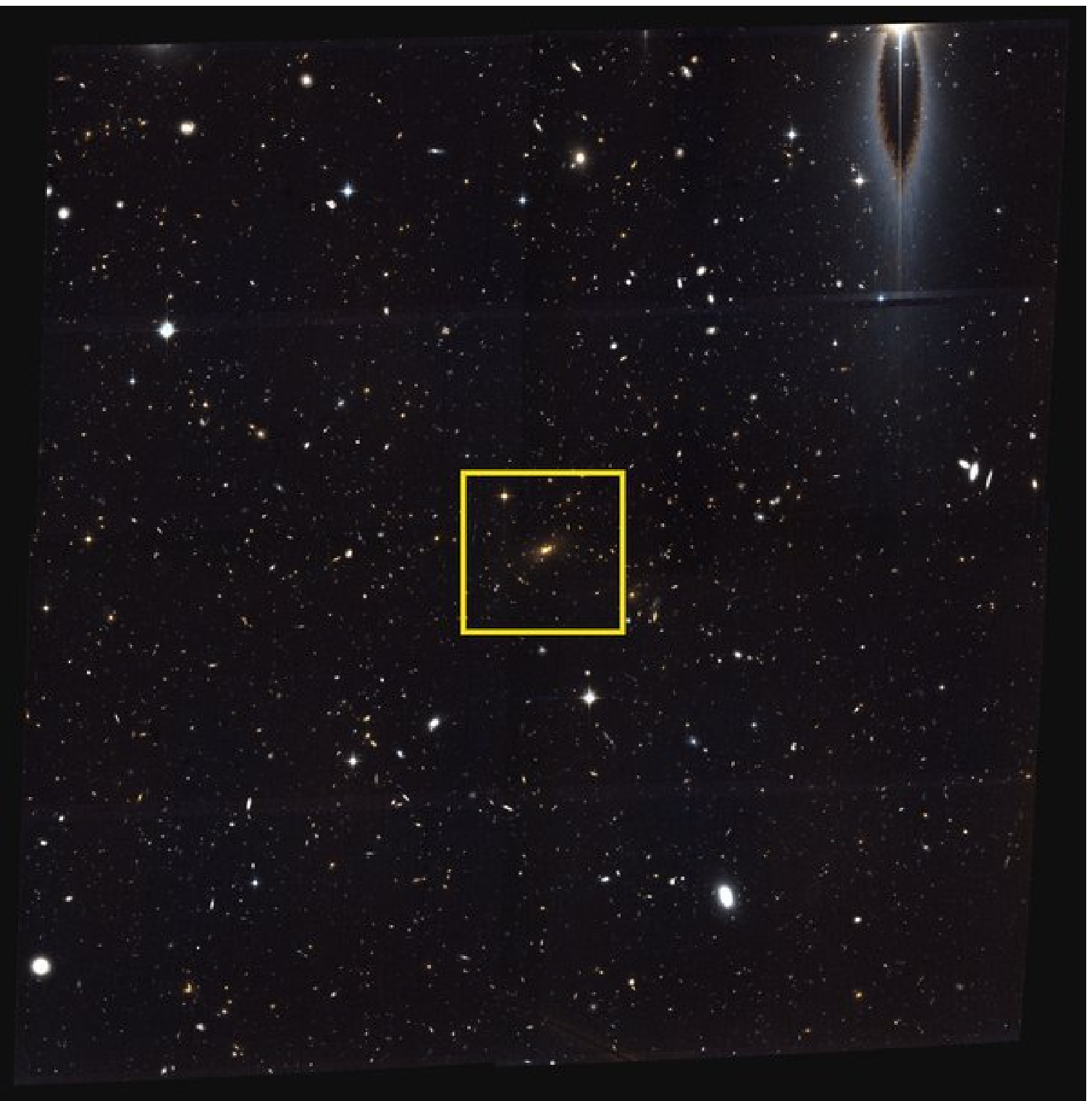}
\caption{Pseudo-color image of CL1226 created from the ACS F606W and F814W images. The entire $\sim6\arcmin\times6\arcmin$ field of
the cluster dithered in a $2\times2$ tiling is displayed. Approximately, north is up and east is left. The ``feather-like" feature
near the northwestern corner is attributed to the internal reflection of the bright star.
The yellow square denotes the central $30\arcsec\times30\arcsec$ region.
\label{fig_cl1226}}
\end{figure}

CL1226 was observed with the Wide Field Camera (WFC) of ACS during April 2004 (PROP ID:9033,
PI:Harald Ebeling).
The cluster was imaged in F606W and F814W (hereafter $v_{606}$ and $i_{814}$, respectively) in a 2$\times$2 mosaic pattern 
covering $\sim6\arcmin\times6\arcmin$
with integrated exposure per pointing of $4000$ s.
We used CALACS (Hack et al. 2003) to perform low level CCD processing and APSIS (Blakeslee et al. 2003) to 
create final mosaic images. A Lanczos3 kernel (windowed sinc function) with a 0.05$\arcsec$ output scale 
was chosen for drizzling (Fruchter and Hook 2002). This combination of drizzling parameters has been
extensively tested in our previous weak-lensing analyses (e.g., Jee et al. 2005a), and provides relatively sharp
point spread functions (PSF) with small noise correlations between pixels. 
The pseudo-color composite of the cluster image is shown in Figure~\ref{fig_cl1226}. We use the $v_{606}$ and
$i_{814}$ images to represent the blue and red intensities, respectively, whereas the mean of the two images is used for the green
intensity. Figure~\ref{fig_cl1226}a displays the entire $\sim6\arcmin\times6\arcmin$ field of the cluster. Approximately, north is up
and east is left. The ``feather-like" feature
near the northwestern corner is attributed to the internal reflection of bright stellar light.
The central $30\arcsec\times30\arcsec$ region marked with a yellow square is magnified in Figure~\ref{fig_cl1226}b.
The cluster red-sequence galaxies are easily identified by their distinct colors. We also observe many blue strongly lensed arc candidates.

We created a separate detection image by weight-averaging the two passband images with their inverse variance maps
and SExtractor (Bertin \& Arnouts 1996) was run in dual-image mode;
objects were detected  
by searching for at least
5 contiguous pixels above 1.5 times sky rms values in the detection image while photometry is performed on each passband image.
We set {\tt CLEAN=Y} with {\tt CLEAN\_PARAM=1.2} to
let spurious detections around the brightest objects be automatically removed.
By visual inspection we identified additional 736 objects that should not be used for weak-lensing. They
include stars, diffraction spikes, merged/fragmented objects, missed cosmic rays, stray light near very bright
objects, etc.
The final catalog contains a total of 11910 objects.

\subsection{Source Galaxy Selection \label{section_source_selection}}

We base our source galaxy selection on the objects' $v_{606}-i_{814}$ colors and $i_{814}$ magnitudes, assuming
that a significant fraction of background galaxies are faint ($i_{814}>24$) and bluer ($v_{606}-i_{814}<1.2$) than the red-sequence of the cluster.
This is a common approach in weak-lensing analyses when limited HST colors are available. 
Although complementary ground-based observations help us to obtain good photometric redshifts for bright
objects [e.g., COSMOS (Scoville et al. 2006) or AEGIS (Davis et al. 2007)], 
most lensing signals (especially for high-redshift clusters such as CL1226) come
from a very faint population, for which it is difficult to obtain reliable, HST PSF-matched colors from the ground. 

\begin{figure}
\plotone{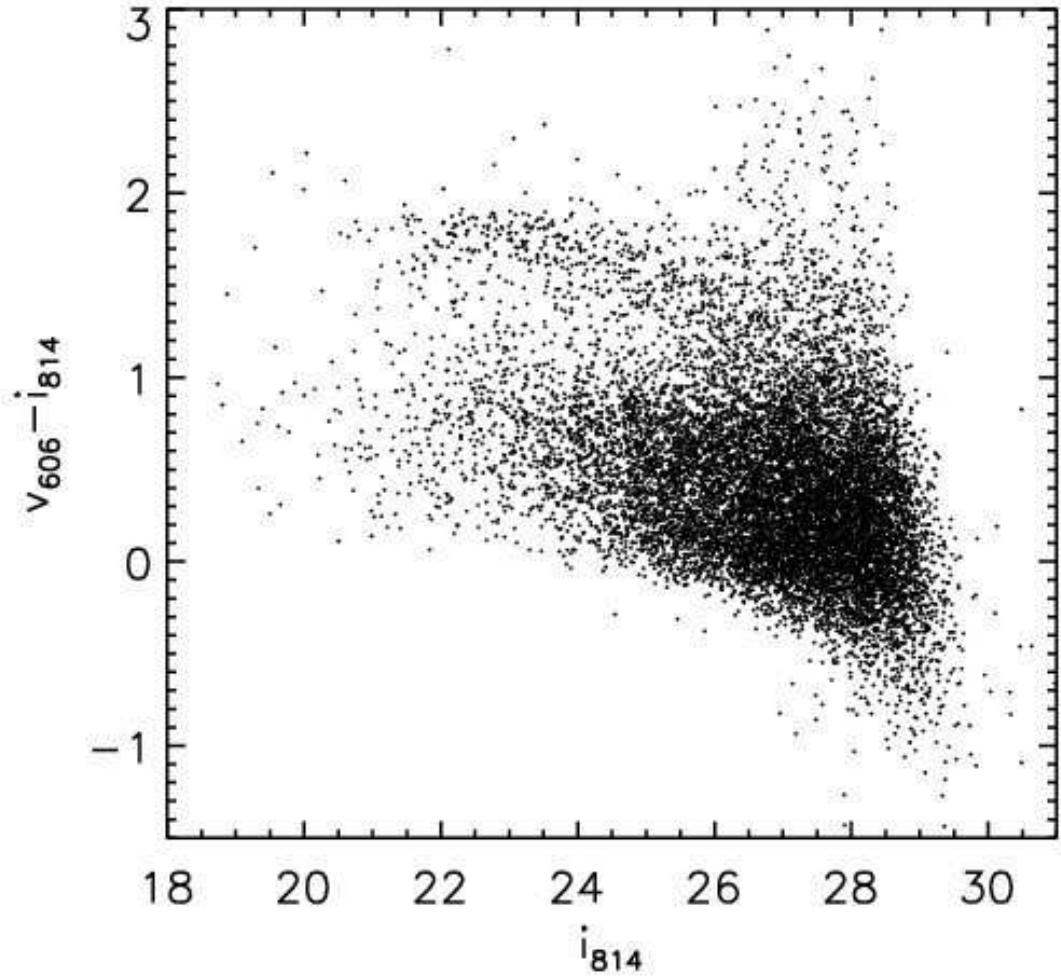}
\caption{Color-magnitude diagram of CL1226. The so-called 4000~\AA~break at $z=0.89$ is redshifted to $\sim7560$~\AA~ and
is nicely bracketed by $v_{606}$ and $i_{814}$ filters. The red-sequence of the cluster is well defined at $v_{606}-i_{814}\sim1.8$
for galaxies brighter than $i_{814}\sim26$.
\label{fig_cmr}}
\end{figure}

\begin{figure}
\plotone{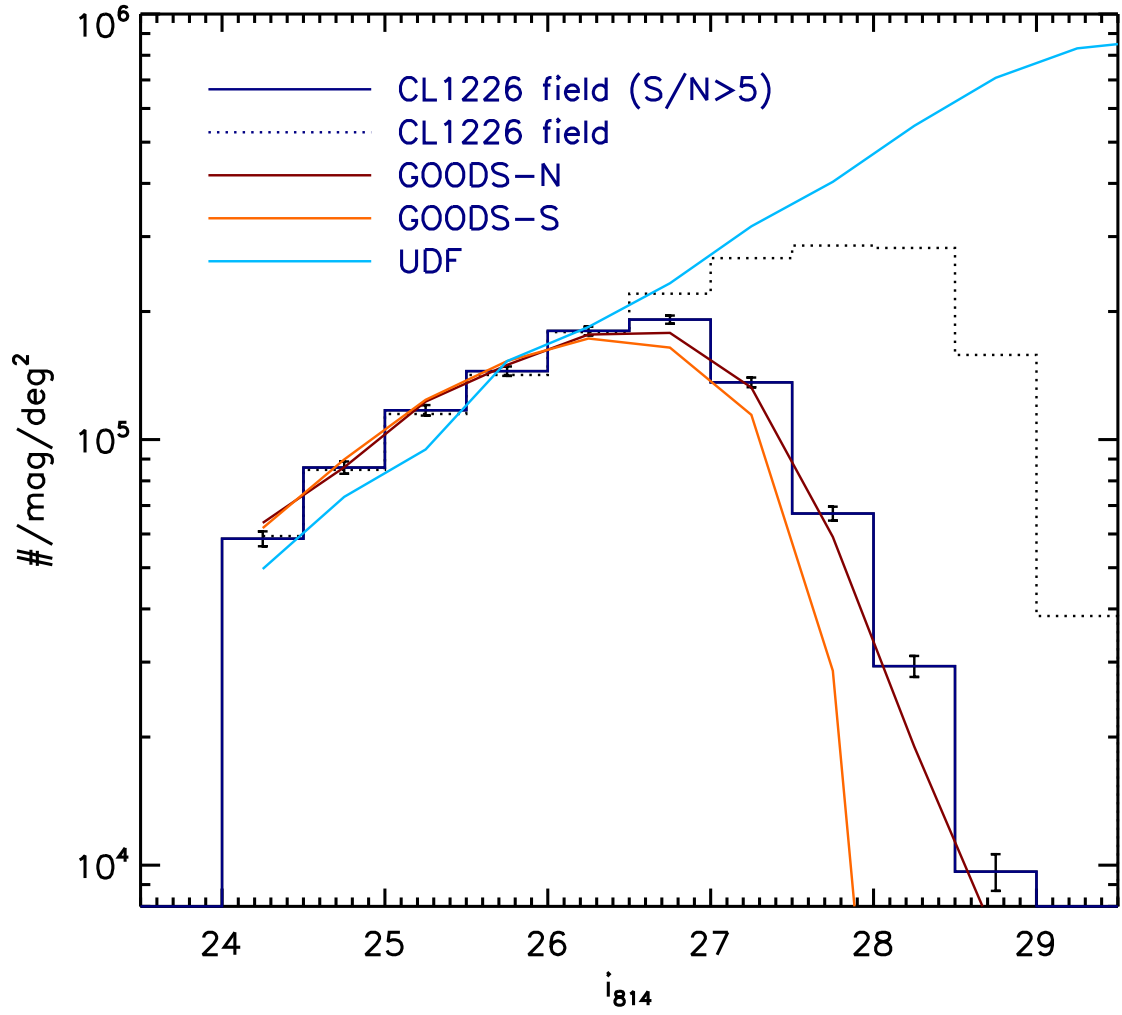}
\caption{Magnitude distribution of source galaxies. We compare the normalized histogram of the source population 
in the cluster field with
those created from the GOODS and UDF data. The comparison shows no indication of excess in the cluster field due to the potential
blue cluster member contamination. 
\label{fig_hist}}
\end{figure}

We show the $i_{814}$ versus $v_{606}-i_{814}$ color magnitude diagram of CL1226 in Figure~\ref{fig_cmr};
we use SExtractor's MAG\_AUTO and MAG\_ISO values 
for $i_{814}$ magnitudes and $v_{606}-i_{814}$ colors, respectively.
The redshifted 4000~\AA~break of galaxies at $z=0.89$ is nicely straddled by the $v_{606}$ and
$i_{814}$ filters. This makes the cluster galaxies clearly visible at $v_{606}-i_{814}\sim1.8$.
down to $i_{814}\sim26$. 
We define our source population as the $24<i_{814}\lesssim28$ and $v_{606}-i_{814}<1.2$
galaxies with ellipticity errors less than 0.2 in one of the two filters (see \textsection\ref{section_ellipticity} for
details on the ellipticity measurement). The resulting number density is $\sim124$ $\mbox{arcmin}^{-2}$ (a total of 4745 objects).
Considering the high-redshift of the cluster, we suspect that a non-negligible
fraction of the cluster members might be bluer than the cluster red-sequence (Butcher \& Oemler 1984). Therefore,
it is important to estimate how much our source catalog might contain blue cluster members despite the clear
presence of the red-sequence. For control fields, we use the Great Observatories Origins Deep Survey (GOODS; Giavalisco et al. 2004)
ACS images and the Ultra Deep Field (UDF; Beckwith 2003) ACS images. Because of the large field, the GOODS data
serves as a fair comparison sample whereas the UDF data provides good statistics of faint galaxies beyond the
detection limit of our cluster observation. The F775W filter (hereafter $i_{775}$) is used in both GOODS and UDF observations
instead of the $i_{814}$ filter. Hence, we transformed $v_{606}-i_{775}$ colors to $v_{606}-i_{814}$ colors to maintain
the consistency.
After selecting galaxies with the same selection criteria ($24<i_{814}\lesssim28$ and $v_{606}-i_{814}<1.2$), we compared
the magnitude distribution of these surveys with that of the cluster observation (Figure~\ref{fig_hist}).
The comparison shows no indication of excess in the cluster field due to the potential blue cluster member contamination. 
This result is consistent with our results in Jee et al. (2005a; 2005b), where we also examined the possible impact of the
blue cluster member contamination to the source catalogs for weak-lensing analyses of the two $z\sim0.83$ clusters.

Ellis et al. (2006) studied the color magnitude relation of CL1226 and reported that
a low fraction ($\sim33$\%) of the 27 spectroscopically confirmed members possess E or S0 early-type morphologies while
some galaxies with late-type morphology lie on the color-magnitude relation defined by the early-type galaxies
(four galaxies within 0.1 mag in the $V-K$ color magnitude relation).
Because the study is based on a small number of bright cluster members ($K>19.6$), it is difficult to apply the result
to faint magnitudes. Our analysis above shows that the fraction of the blue cluster members is negligibly small
at least for the population selected by the ($24<i_{814}\lesssim28$ and $v_{606}-i_{814}<1.2$) criteria.

Now with the source catalog at hand, we need to estimate the redshift distribution in order to put our subsequent lensing
analyses on the proper scale. We utilize the publicly available photometric redshift catalog of the UDF (Coe et al. 2006).
The UDF covers a small area (though twice as large as the Hubble Deep Field), but thanks to the unprecedented depth of HST-based observations in
the $B_{435}$, $V_{555}$, $i_{775}$, $z_{850}$, $j_{110}$, and $h_{160}$ filters, it provides
high-fidelity photometric redshifts for the galaxies beyond the limiting magnitudes of the CL1226 data.
To account for different depth between the cluster field and UDF, we estimate the redshift distribution per each magnitude bin
and correct for the difference in the normalized number density. The resulting mean redshift of the source population
is determined to be $\bar{z}=1.71$. Of course, this value should not be confused with the {\it effective} redshift
of the source population because objects at redshifts smaller than the cluster redshift dilute the signal.
In fact, in weak-lensing studies this lensing efficiency is expressed in terms of $\beta$:
\begin{equation}
\beta =  \mbox{max} \left ( 0, \frac{D_{ls}} {D_s} \right ). \label{eqn_beta}
\end{equation}
\noindent
where $D_s$ and $D_{ls}$ are the angular diameter distance from the observer to the source and 
from the lens to the source, respectively.
We obtain $<\beta>=0.265$ for the given cosmology, which corresponds to $z_{eff}=1.373$.
Another important quantity that affects our subsequent lensing analysis is the width of the redshift
distribution, which is often expressed in terms of $<\beta^2>$. Seitz \& Schneider (1997) found that
under a single redshift source plane assumption the measured shear $g\prime$ is
overestimated by
\begin{equation}
g\prime = \left[1 + (\frac{<\beta^2>}{<\beta>^2}-1) \kappa \right]g
\end{equation}
\noindent
For the current source population, we obtain $<\beta^2>=0.12$ and therefore the measured shear
is overestimated by $(1+0.71\kappa)$. This correction becomes increasingly important with 
lens redshift and should be included in high-redshift cluster lensing analyses.

\subsection{PSF Modeling and Ellipticity Measurements \label{section_ellipticity}}

Weak-lensing measures a subtle distortion of background galaxy images and therefore it is
important to remove any instrumental effect, which can mimic gravitational lensing signals.
When geometric distortion and image registration are done carefully, the most
important remaining task is PSF modeling. Because anisotropic PSF can induce
a false lensing signal and the impact becomes greater for fainter galaxies,
which contain  more signal, a great amount of efforts and time is 
invested on studying the PSF of any instrument before a signal is extracted.

Although the ACS PSF is far smaller than what one can achieve from the ground, 
it still measurably affects the shapes of objects whose sizes at the surface brightness limit are comparable
to the PSF. This places great importance on deep imaging (larger size at low surface brightness)
and good understanding of the PSF variations.
It has been known that ACS PSFs vary in a complex way with time and 
position (Krist 2003; Jee et al. 2005a; Sirianni et al. 2005). In Jee et al. (2007b),
we presented a principal component analysis of the ACS PSF 
and made
the ACS PSF library publicly available based on archival ACS images of stellar fields.
In this work, we use the PSF library of Jee et al. (2007b) to model the PSF variation
in the CL1226 field. Those who are interested in the method in detail are referred
to the paper. Below we briefly describe the procedure and the result specific
for our cluster analysis.

\begin{figure}
\plotone{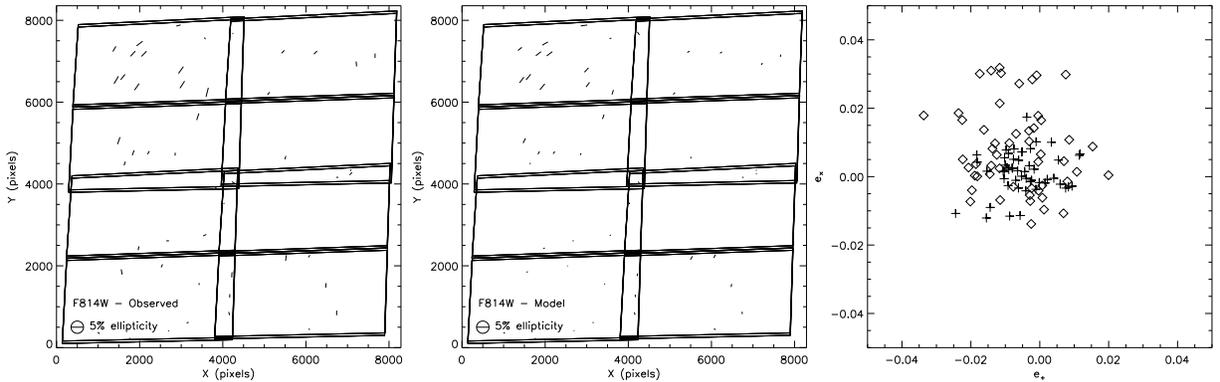}
\caption{Comparison of ellipticity between the observed and the modeled PSFs in the mosaic $i_{814}$ image of the cluster. 
The outlines of the ACS pointings are displayed in the observed orientation (north is down and east is right).
We represent
the magnitude and the direction of ellipticity with the length and the orientation of the sticks, respectively.
The encircled stick in the lower left corner shows the size of the $(a-b)/(a+b)=0.05$ ellipticity.
We first determined a matching template from the PSF Library for each exposure and then applied the same shift and rotation
to align the template to the final mosaic image as that used in our image registration. The final PSFs are the results of stacking all the contributing PSFs.
The PSF models (middle) obtained in this way closely resemble the
observed pattern (left). The $e_{+}$ versus $e_{\times}$ plot (right) shows that both the centroid and the dispersion
of the observed points (diamond) are significantly improved in the residuals (`+' symbol).
\label{fig_psf}}
\end{figure}

We first derive a PSF model for an individual exposure and then shift/rotate the model PSFs
with respect to the final mosaic image in a similar way to our  
image registration procedure. In each exposure, there are typically 8-15 high S/N stars available,
which can be used to find the best-matching template from the library. 
The final PSFs are the results of stacking all the contributing PSFs.
In Figure~\ref{fig_psf} we show the comparison of the observed $i_{814}$ PSFs with the modeled ones
(similar results are obtained for the $v_{606}$ PSFs). 
The PSF model ($middle$) obtained in this way closely resembles the
observed pattern ($left$). The $e_{+}$ versus $e_{\times}$ plot (right) shows that both the centroid $<\bvec{e}>\simeq(-7\times10^{-3},8\times10^{-3})$ 
and the dispersion $<|\bvec{e}|^2>^{0.5}\simeq0.02$
of the observed points (diamond) are significantly improved in the residuals (`+' symbol).
The centroid and dispersion of the residuals are $<\delta \bvec{e}>\simeq(-5\times10^{-3},1\times10^{-3})$ and 
$<|\delta \bvec{e}|^2>^{0.5}\simeq0.01$, respectively.

We determine object ellipticities by fitting a PSF-convolved elliptical Gaussian
to the images. 
In theory, this is equivalent to the method proposed by Bernstein \& Jarvis (2002)
although the implementation is different. Instead of fitting an elliptical Gaussian
to an object, they shear the object progressively until it fits a $circular$ Gaussian.
This scheme is conveniently implemented by first decomposing galaxy shapes with shapelets (Bernstein \& Jarvis 2002; Refregier 2003)
and then by applying shear operators to the shapelet coefficients until the object's quadrupole moments disappear.
We adopt the method in our previous analysis 
(Jee et al. 2005a; 2005b; 2006). We noted in Jee et al. (2007a) however that
directly fitting an elliptical Gaussian to the pixelized object reduces aliasing compared
to the shapelet formalism, particularly when the object has extended features. The issue
was important in Jee et al. (2007a) because the lens was a low-redshift strong-lensing
cluster. The ellipticities of the strongly lensed arc(let)s were substantially
underestimated if the shapelet formalism is employed.
Although the CL1226 cluster images do not show such a large number of
arc(let)s, we use the same pipeline of Jee et al. (2007a) in the current analysis.
Fitting a PSF-convolved elliptical Gaussian to pixelated images
is more numerically stable for faint objects, and also provides straight-forward error estimates 
in the results.
Because we have two passband images available, the finalcombined  ellipticities of objects are
given as weighted averages. 

\begin{figure}
\plottwo{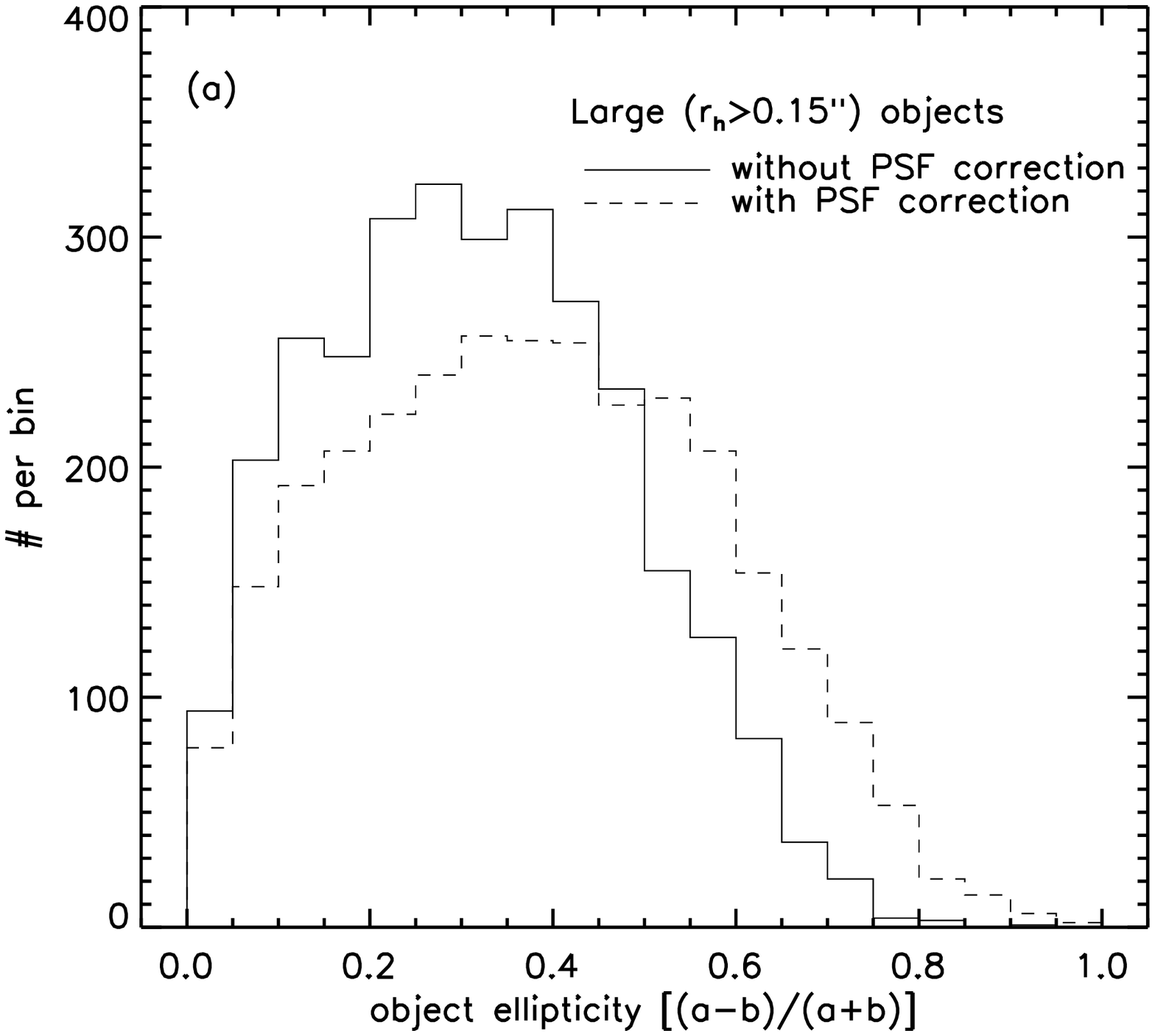}{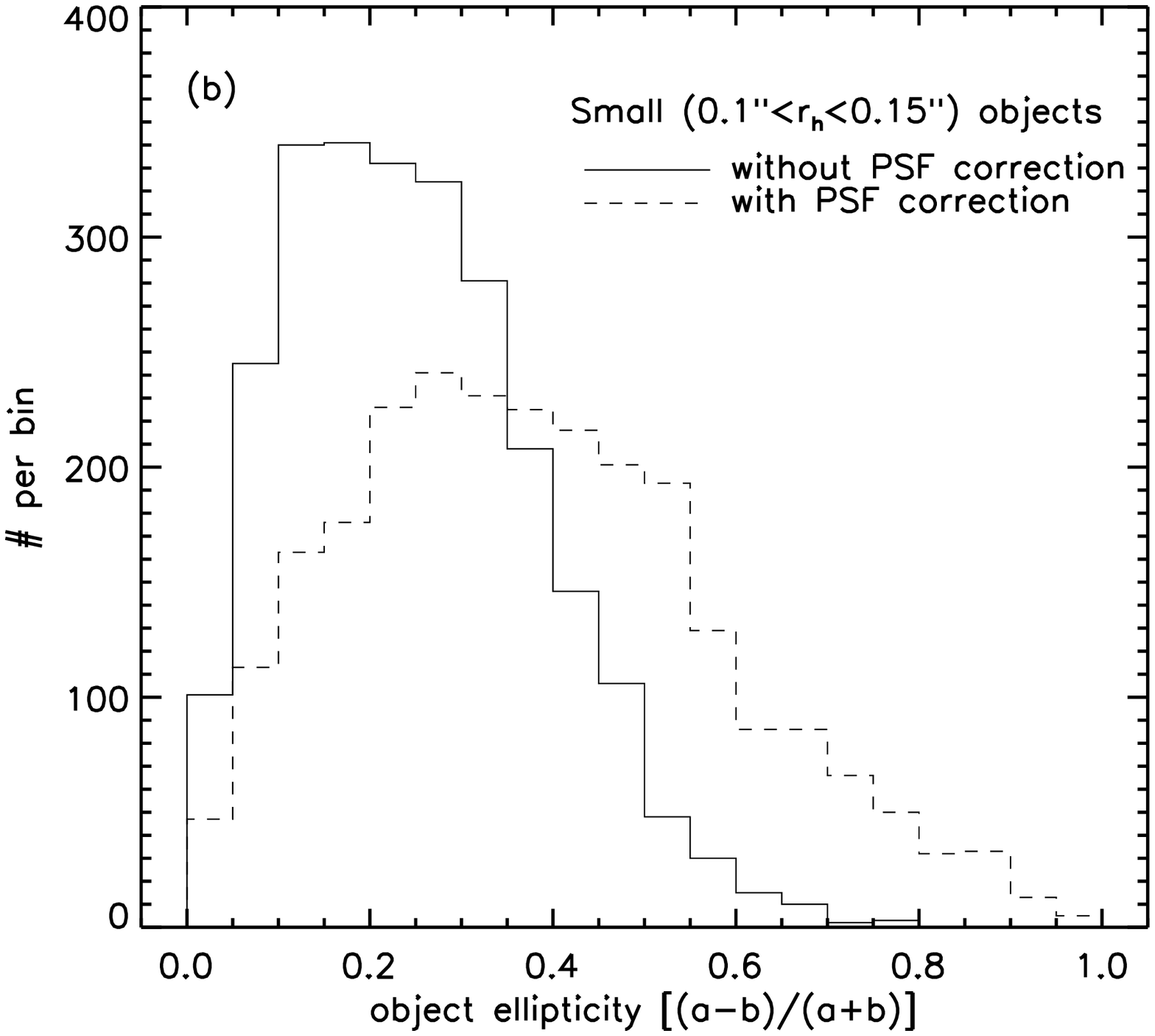}
\caption{Ellipticity distribution of non-stellar objects in the CL1226 field. We only include the objects with a S/N$\sim5$ or greater
at least in one passband. The ellipticity distribution of $r_h > 0.15\arcsec$ objects is slightly affected after
PSF correction (a) whereas the change is significant for smaller ($0.1\arcsec<r_h < 0.15\arcsec$) objects (b). 
\label{fig_e_distribution}}
\end{figure}

Figure~\ref{fig_e_distribution} displays the ellipticity distribution 
of non-stellar objects in the cluster field. We only include the objects with a S/N$\sim5$ or greater
at least in one passband. The ellipticity distribution of the $r_h$\footnote{$r_h$ represents a half light radius.} $> 0.15\arcsec$ objects is slightly affected after
PSF correction (Figure~\ref{fig_e_distribution}a) whereas the change is significant for smaller ($0.1\arcsec<r_h < 0.15\arcsec$) objects (Figure~\ref{fig_e_distribution}b). 
Note that much of the useful lensing signal comes from this ``small'' galaxy population (only slightly larger than the instrument PSFs 
$r_h\sim0.06\arcsec$)
for high-redshift clusters.
This illustrates that
that, even if ACS PSFs are small, one must carefully account for their effects to maximize the full resolving power of ACS.

A potentially important factor in determining object shapes in addition to the PSF correction discussed above is 
the charge transfer efficiency (CTE) degradation of ACS. Riess and Mack (2004) reported strong evidence 
for photometric losses in the parallel direction for ACS.
Rhodes et al. (2007) studied the CTE-induced charge elongation in the context of weak-lensing studies and established
an empirical prescription, where the strength of the elongation is proportional to 1) the distance from the read-out register,
2) the observation time, and 3) the inverse of the S/N of objects. When we assume that equation 10 of Rhodes et al. (2007) is
also applicable to the current data, we obtain 
$\delta e_{+}\sim0.01$ for the faintest object that is also farthest from the read-out register; on average however for all sources 
throughout the entire field
the required correction is $\delta e_{+}\sim0.003$. Hence, the CTE-induced elongation is much smaller than
the weak-lensing signal of the cluster and the statistical noise (object ellipticity dispersion)
to the extent that the effect can be safely ignored in our subsequent analysis. Heymans et al. (2008)
obtained a similar result in their weak-lensing study of the Abell 901/902 supercluster whose ACS data were taken about 1 year
after CL1226 was observed. 
We are also investigating the CTE-induced elongation issue independently by analyzing
cosmic-rays. Because cosmic-rays are not affected by the instrument PSF, their study
enables us to nicely separate the PSF effect from the CTE degradation effect. Our preliminary
result is consistent with that of Rhodes et al. (2007) in the sense that the required correction
is negligibly small for the current observation.
We will present our result of the CTE-induced elongation study elsewhere
along with our future publication of the weak-lensing analysis of the $z\sim1.4$ cluster XMMU J2235.3-2557 (M. Jee et al. in preparation), whose
ACS data were taken in the 2005-2006 years and thus are potentially subject to greater bias due to the CTE degradation.

\section{WEAK-LENSING ANALYSIS}
\subsection{Two-Dimensional Mass Reconstruction \label{section_mass_map}}

\begin{figure}
\plotone{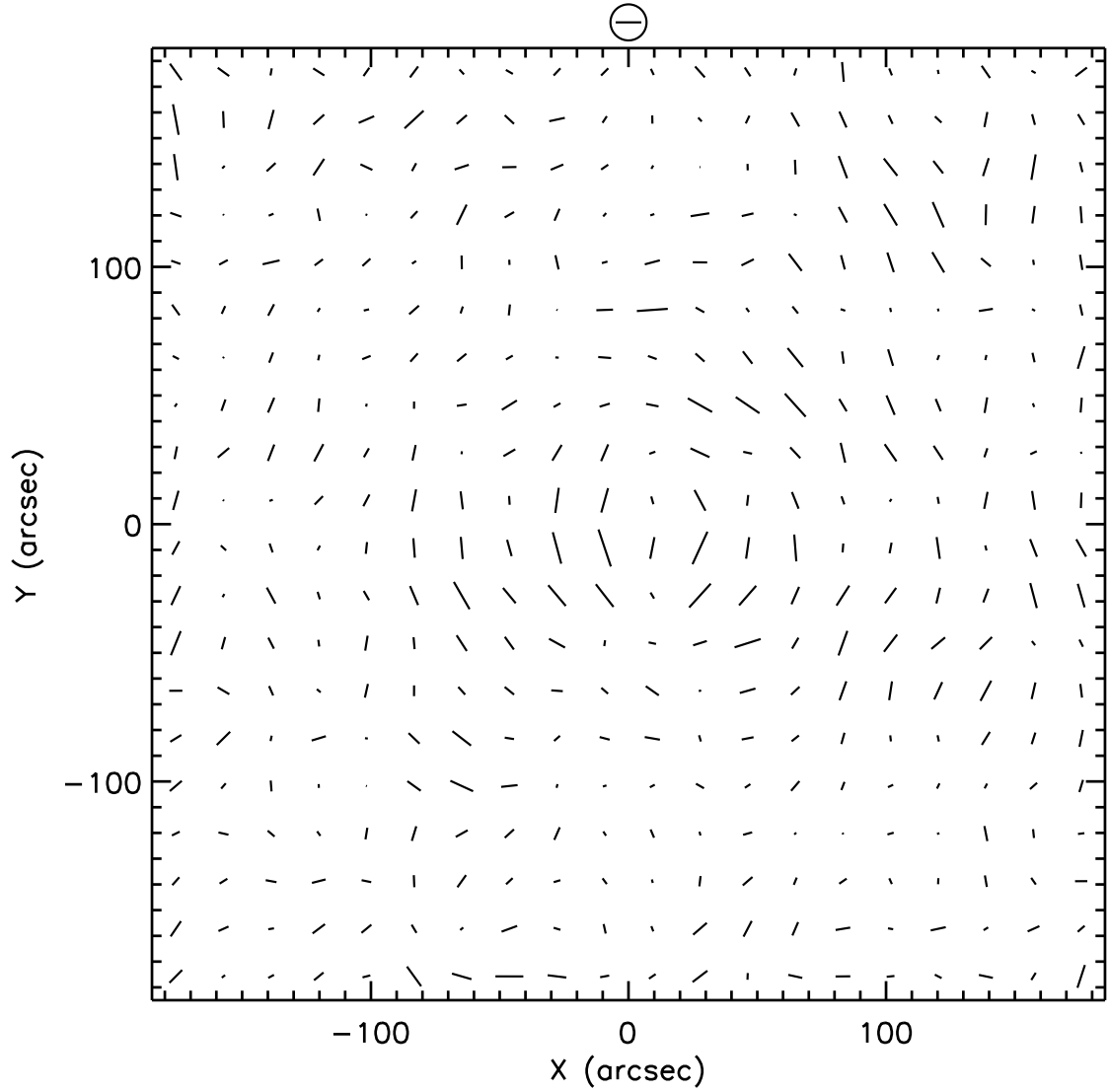}
\caption{Ellipticity distribution of source galaxies. We smooth the source galaxy ellipticity with a FWHM=$20\arcsec$ Gaussian kernel.
The encircled stick above the plot shows the size of the $g=0.1$ shear. The tangential alignment around the center of the cluster is clear.
\label{fig_whisker}}
\end{figure}

One of the easiest ways to visually identify the presence of a lensing signal is to plot a smoothed two-dimensional distribution
of the source galaxies' ellipticity. Massive clusters shear shapes of background galaxies in such a way that they appear on average tangentially
aligned toward the center of the clusters. In Figure~\ref{fig_whisker} we present this so-called ``whisker'' plot obtained by smoothing
the source galaxy ellipticity map with a FWHM=$20\arcsec$ Gaussian kernel. As in the case of Figure~\ref{fig_psf}, the length and the orientation
of the sticks represent the magnitude and the direction of the weighted mean ellipticity, respectively. An ellipticity with a magnitude of
$g=0.1$ is shown at the top with a circle for comparison. The tangential alignment around the center of the cluster (the location
of the BCG) is clear.

Many algorithms exist for the conversion of this ellipticity map to the mass density map of the cluster. Due to its simplicity, the classic method of 
Kaiser \& Squires (1993; hereafter KS93) or the real-space version (Fischer \& Tyson 1997)
still is widely used. The KS93
method is based on the notion that the measured shear $\gamma$ is related to the dimensionless mass density $\kappa$ by the
following convolution:
\begin{equation}
\kappa (\bvec{x}) = \frac{1}{\pi} \int D^*(\bvec{x}-\bvec{x}^\prime) \gamma (\bvec{x}^\prime) d^2 \bvec{x} \label{k_of_gamma}. 
\end{equation}
\noindent
where $D^*(\bvec{x} )$ is the complex conjugate of the convolution kernel $D(\bvec{x} ) = - 1/ (x_1 - i x_2 )^2$. The method assumes that the shear $\gamma$ is
directly measurable whereas in fact it is the reduced shear $g=\gamma/(1-\kappa)$ that we can measure directly. Obviously, in the region where
$\kappa$ is small, the assumption is valid. However, near cluster centers, $g$ is often much greater than $\gamma$, and this leads
to overestimation of $\kappa$. In addition, it is not straightforward to incorporate measurement errors 
(i.e., ellipticity errors and shear uncertainties) or priors
in the KS93 scheme. Because measured shears in cluster outskirts have much lower significance, the algorithm 
frequently produces various noise peaks when the smoothing scheme is optimized to reveal significant structures in the cluster center.

These pitfalls are overcome in the new methods such as Marshall et al. (2002) and Seitz et al. (1998), where individual galaxy shapes (not averaged shears)
are used and the resulting mass map is regularized.
In the current paper, we used the mass reconstruction code of Jee et al. (2007a), who modified
the method of Seitz et al. (1998) so that strong-lensing constraints are incorporated. For the current mass reconstruction of CL1226, however, we turned off the
strong-lensing capability of the software and utilized only the weak-lensing data. 

\begin{figure}
\plottwo{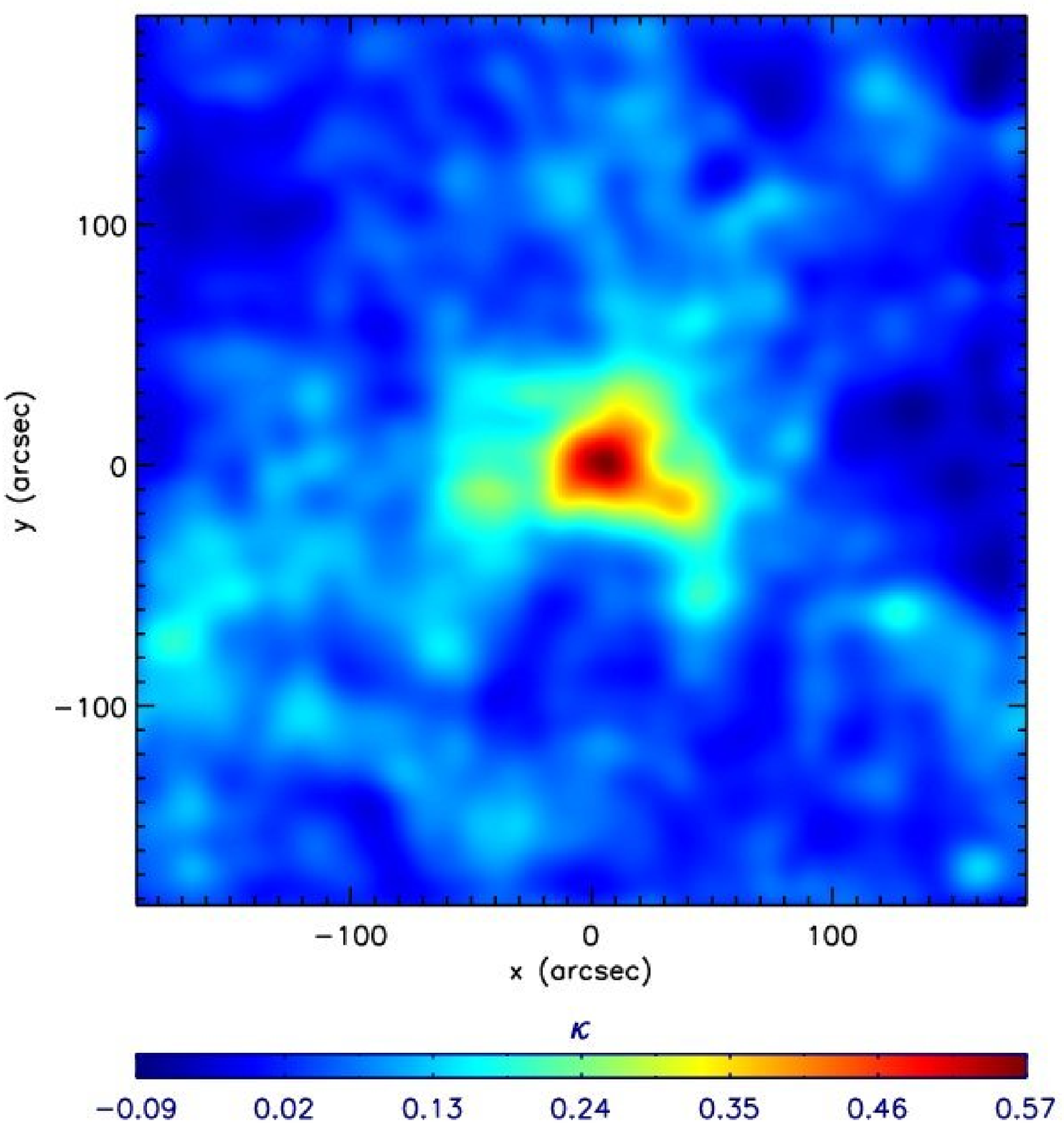}{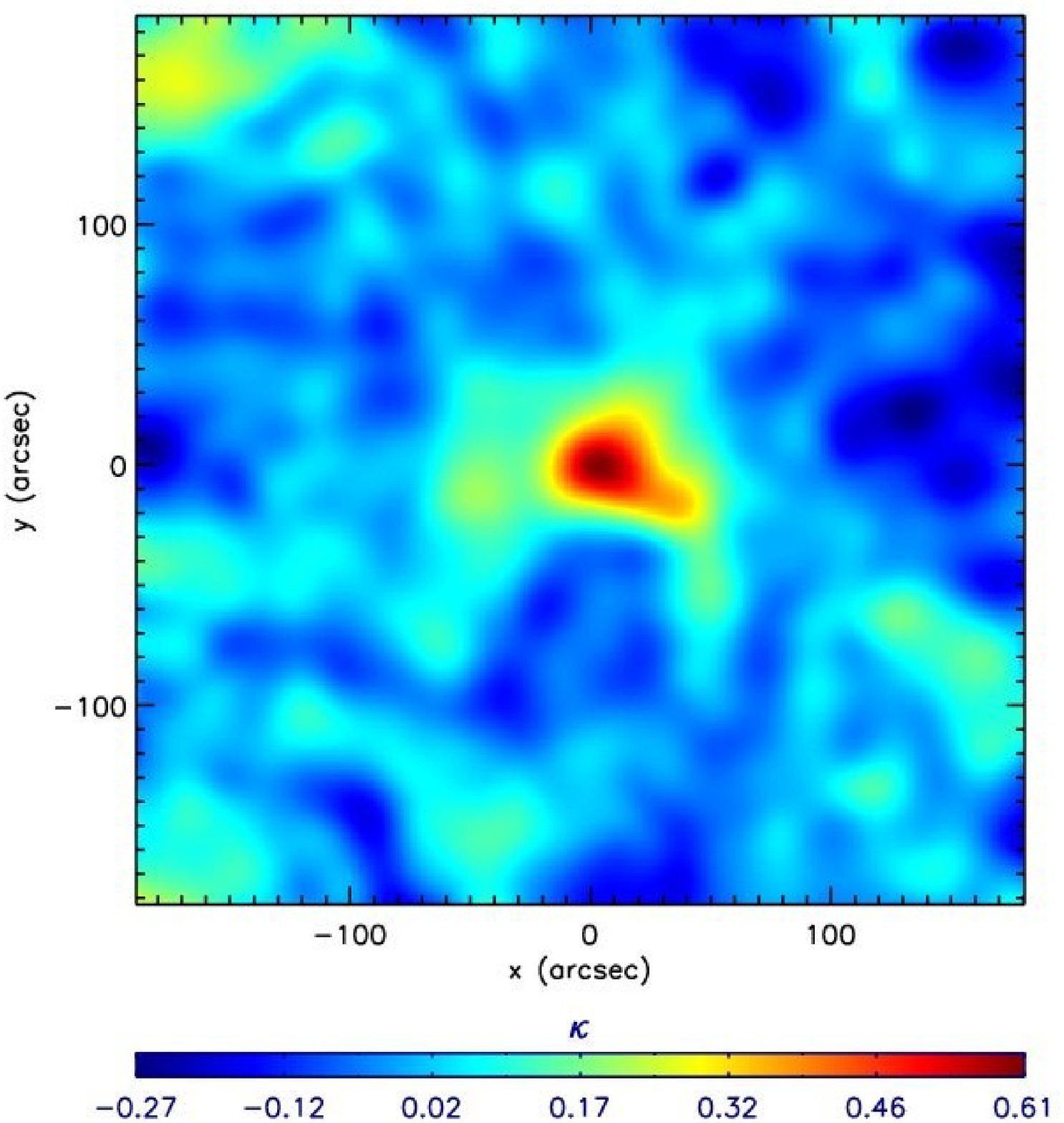}
\caption{Mass reconstruction of CL1226. We used the maximum entropy method to obtain the convergence $\kappa$ of the field ($left$).
The mass-sheet degeneracy ($\kappa \rightarrow \lambda \kappa + 1 -\lambda$) is lifted by constraining $\bar{\kappa}(150\arcsec<r<200\arcsec)$
to be the same as the value given by the NFW fitting (\textsection\ref{section_mass_estimation}).
For comparison, we also show the result created by the KS93 method ($right$). Both maps clearly show the dark matter
concentration near the cluster center. However, in the relatively low $\kappa$ region the KS93 algorithm generates many spurious substructures, which
are the results from both the inadequate (too small kernel) smoothing and the $g\sim\gamma$ approximation.
\label{fig_mass_reconstruction}}
\end{figure}

We present our maximum-entropy mass reconstruction of CL1226 in the left panel of Figure~\ref{fig_mass_reconstruction}. For comparison, we also
display the result obtained by the conventional KS93 algorithm in the right panel. For the KS93 method we choose
a smoothing scale of FWHM$\sim24\arcsec$. The regularization parameter of Jee et al. (2007a) was adjusted in such a way that the result
matches the resolution of the KS93 version at $r\lesssim50\arcsec$.
Both mass reconstructions clearly reveal
the strong dark matter concentration in the cluster center. However, in the relatively low $\kappa$ region the KS93 algorithm produces many
spurious substructures, most of which do not stand out in the maximum-entropy reconstruction.
It is certain that both the inadequate (too small kernel) smoothing and the $g\sim\gamma$ approximation of the KS93 algorithm are the causes
of the artifacts. Therefore, our interpretation hereafter is based on the result from our maximum-entropy reconstruction.

On a large scale the $\kappa$ field shows that the cluster does not possess any significant substructure.  
This relaxed appearance is also indicated by the X-ray emission
from the cluster (Maughan et al. 2007). However, this symmetric mass distribution is somewhat unusual for a cluster at such a high redshift.
In our current hierarchical structure formation paradigm, relaxed clusters are thought to be rare at $z=0.89$, when the universe is at
less than half its current age. Our previous weak-lensing analysis of CL0152 and MS1054 (both at $z\sim0.83$)
revealed significant substructures composed of several mass clumps suggestive of the active formation of the systems. 

\begin{figure}
\begin{center}
\includegraphics[width=12cm]{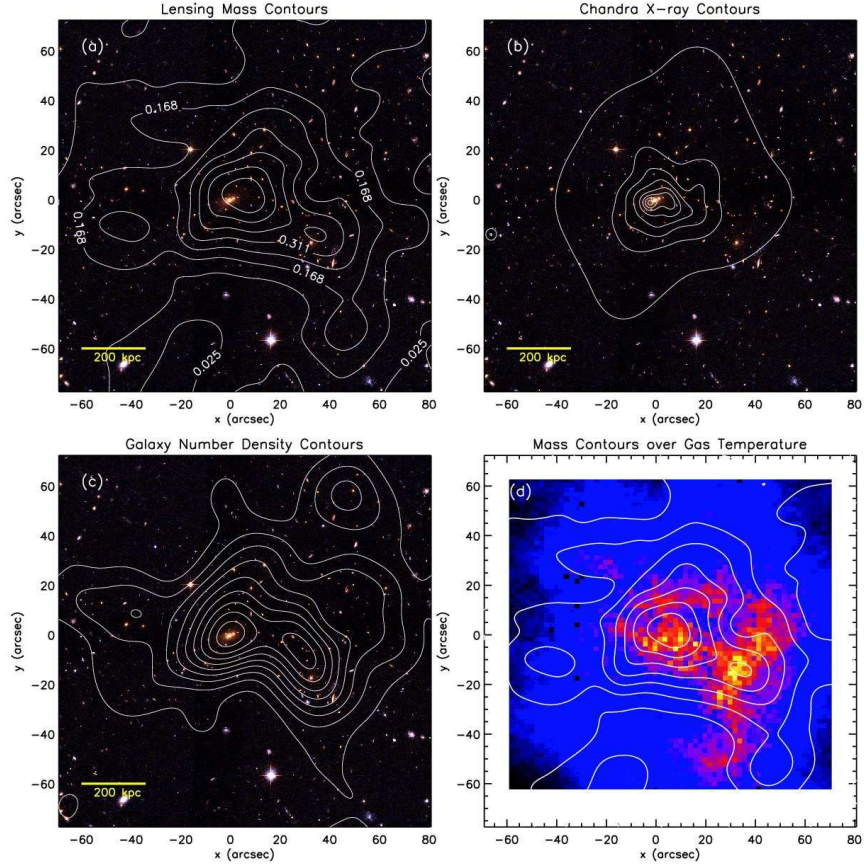}
\caption{Mass, X-ray, and galaxy number density contours in the central $150\arcsec\times150\arcsec$ region of CL1226.
(a) Our weak-lensing mass reconstruction resolves the core substructure, which consists of the dominant clump near the BCG and
the less massive, but distinct, clump to the southwest. (b) The presence of this secondary mass peak is hard to identify
in the adaptively smoothed (Ebeling 2005) $CHANDRA$ X-ray contours although the slight elongation of the X-ray peak toward the secondary
mass peak is marginally suggestive of this feature. (c) The number density contours of the red-sequence candidates (smoothed with a FWHM$=20\arcsec$ Gaussian) show
that the galaxy distribution is similar to the mass distribution.
(d) Mass contours overlaid on the temperature map of Maughan et al. (2007). The alignment is approximate. 
The gas temperature in the region where the southwestern mass clump is detected is unusually high ($12\sim18$ keV).
\label{fig_massxraynum}}
\end{center}
\end{figure}

On a small scale, however, our mass reconstruction resolves the core into two mass clumps, which are separated by $\sim40\arcsec$
(Figure~\ref{fig_massxraynum}a).
Comparing the projected masses within $r<20\arcsec$, 
we estimate that the mass ratio of the two substructures
is approximately 3:2. The more massive mass clump [$M(r<20\arcsec)=(1.3\pm0.1)\times10^{14}~M_{\sun}$] is located near the brightest cluster galaxy, which is also close to the
center of the X-ray peak (Figure~\ref{fig_massxraynum}b). The less massive structure [$M(r<20\arcsec)=(8.5\pm0.6)\times10^{13}~M_{\sun}$] $\sim40\arcsec$ to the southwest is however not
detected in the X-ray surface brightness although we note that the contours near the X-ray peak are slightly elongated toward this secondary mass clump.
Because the mass of this structure is significant and comparable to that of the western mass peak (the most massive among the three) of MS1054,
the apparent absence of the gas overdensity associated with the structure is counter-intuitive.
The weak-lensing mass structure is nevertheless highly consistent with the cluster red-sequence distribution.
We display the number density contours of the cluster red-sequence in Figure~\ref{fig_massxraynum}c. The secondary mass clump
is in good spatial agreement with the cluster red-sequence. Interestingly, Maughan et al. (2007) found that the gas temperature
of the region is much higher ($12-18$ keV) than those of the rest [see Figure~\ref{fig_massxraynum}d where we overplot
the mass contours on top of the temperature map of Maughan et al. (2007)]. They suggested that this temperature structure might relate
to the possible on-going merger indicated by the cluster galaxy distribution and we agree. Our detailed discussion on the comparison
between the mass, X-ray intensity, gas temperature, and galaxy distributions is deferred to \textsection\ref{section_merger}.

\subsection{Tangential Shear, Cosmic Shear Effect and Mass Estimation \label{section_mass_estimation} }

In addition to the two-dimensional mass reconstruction discussed previously, 
tangential shear is also a useful measure of total lensing mass.
By taking azimuthal averages, one can lower the effect of shot noise
and more easily determine the presence of the lensing than in the two-dimensional analysis particularly when the
signal is weak. Therefore, many authors prefer to use tangential shear profiles in the estimation of cluster masses
assuming an azimuthal symmetry. In the current paper, we present the tangential shear profile of CL1226 first, and
then estimate the mass based on the profile. Finally, we also compute the mass using the two-dimensional mass map and
compare the results.

\begin{figure}
\plotone{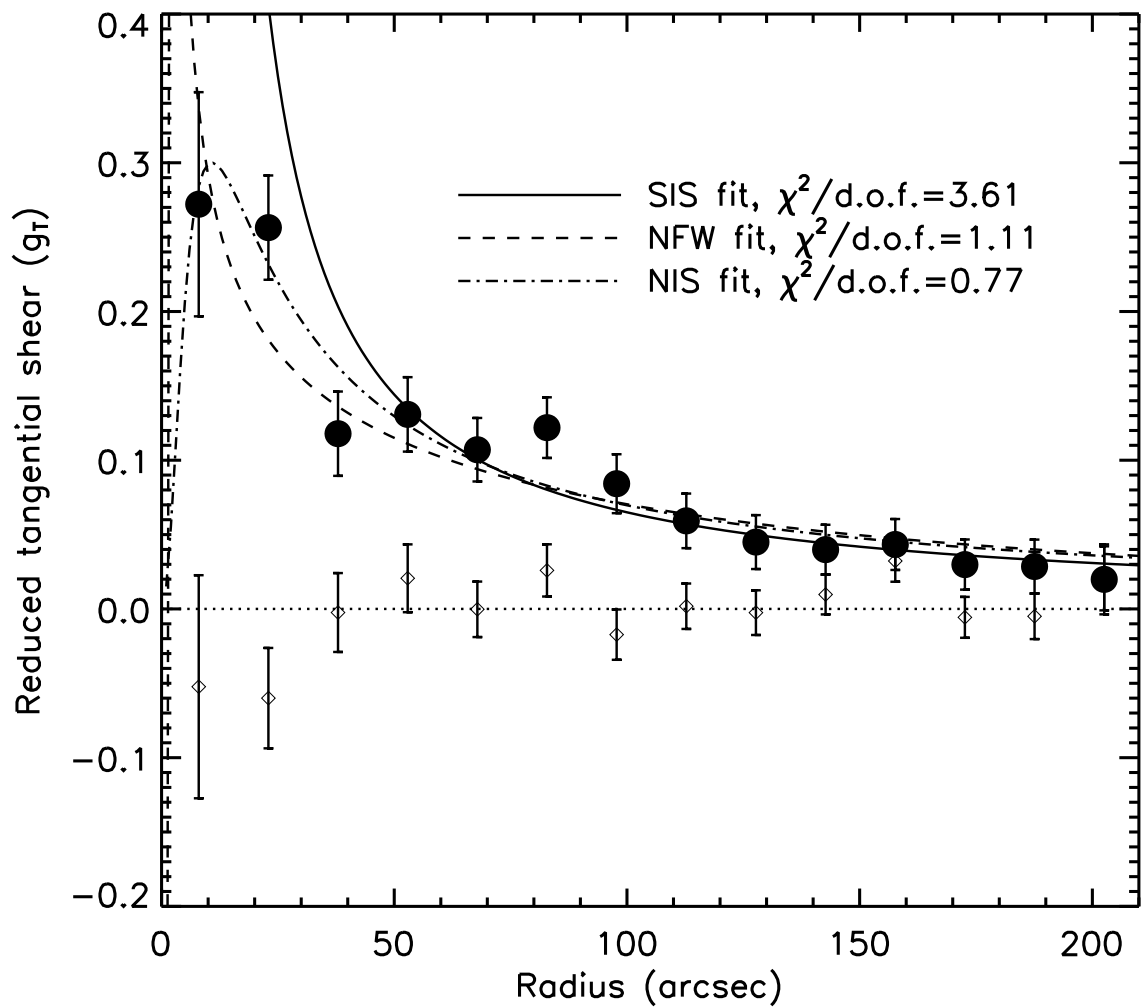}
\caption{Tangential shear measured around CL1226. The filled circles represent the tangential shears. Overplotted are
the fitting results with different parametrized models. Open diamond symbols show the 45$\degr$ rotation test result (also often referred to as a B-mode test).
\label{fig_tan_shear}}
\end{figure}

Tangential shear is defined as
\begin{equation}
 g_T  = < -  g_1 \cos 2\phi - g_2 \sin 2\phi > \label{tan_shear},
\end{equation}
\noindent
where $\phi$ is the position angle of the object with respect to the lens center, and $g_{1(2)}$ is
a reduced shear $g_{1(2)}=\gamma_{1(2)} /( 1- \kappa )$; true shears $\gamma_1$ and $\gamma_2$
are related to the lensing potential $\psi$ by $\gamma_1=0.5 (\psi_{11} -\psi_{22})$
and $\gamma_2=\psi_{12} =\psi_{21}$.
If no lensing is present, the reduced tangential
shear $g_T$ must be consistent with zero. The filled circles in Figure~\ref{fig_tan_shear}
show the reduced tangential shears of CL1226 centered on the BCG. The lensing signal from the cluster
is clear out to the field limit; we note that at $r>180\arcsec$ we
cannot complete a circle. The observed reduced shears at $r>50\arcsec$ decrease monotonically with radius.
This is in accordance with the mass map of the cluster discussed in \textsection~\ref{section_mass_map}, which
shows no major asymmetric substructures outside the main clump.
The diamond symbols represent our measurement of tangential shear when the background galaxies are rotated by 45$\degr$.
This test shows that the B-mode signal is consistent with zero as expected.

\begin{figure}
\plotone{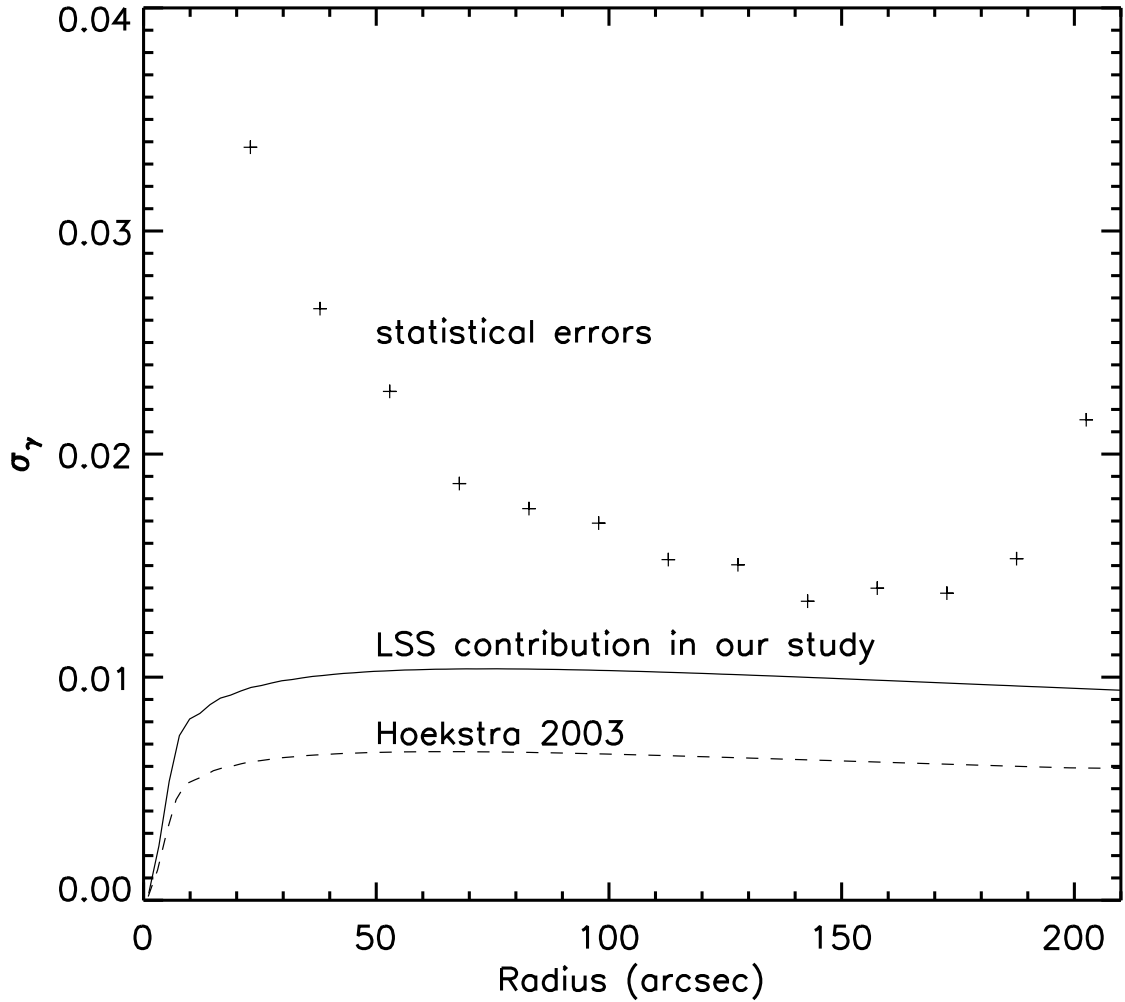}
\caption{Background structure effects on tangential shear measurements. 
We followed the method of Hoekstra (2003) to evaluate the effect of the cosmic shear
on our tangential shear measurements for the current source population (solid). For comparison, we reproduce here the prediction of Hoekstra (2003)
for their $20<R<26$ ($\bar{z}=1.08$) sample (dashed). Note that the effect is substantially ($\sim50$\%) higher in our sample because the mean redshift of
the source population is also higher ($\bar{z}=1.71$). However, in the $r\lesssim200\arcsec$ region, where
we measure our tangential shears for CL1226, the errors induced by the cosmic shear are still lower than the
statistical errors (`+' symbols).
\label{fig_cs_effect}}
\end{figure}

The error bars in Figure~\ref{fig_tan_shear} include only the statistical uncertainties determined by
the finite number of source galaxies in each radial bin. Hoekstra (2003) demonstrated that background
structures (cosmic shear effects) are important sources of uncertainties and need to be considered
in cluster mass estimation. Following the formalism of Hoekstra (2003), we evaluated the
background structure effect on the uncertainties in the tangential shear measurements. The solid
line in Figure~\ref{fig_cs_effect} displays the predicted errors $\sigma_{\gamma}$ for the redshift
distribution of the source population in the current cosmology. For comparison, we approximately reproduce here the prediction of Hoekstra (2003)
for their $20<R<26$ sample ($\bar{z}=1.08$, dashed). 
Note that the cosmic shear effect is substantially ($\sim50$\%) higher in our sample because the mean redshift of
our source population is also significantly higher ($\bar{z}=1.71$). However, in the $r\lesssim200\arcsec$ region, where
we measure our tangential shears for CL1226, the errors induced by the cosmic shear are still lower than the
statistical errors (`+' symbol) and thus the effect is minor in our cluster mass estimation
Nevertheless, in the following analysis, we include this
cosmic shear effect in the quoted errors.

We characterize the reduced tangential shears with three parametric models: singular isothermal sphere (SIS), 
Navarro-Frenk-White (NFW; Navarro et al. 1997), and non-singular isothermal sphere (NIS).
In these parametric fits,  we use all the data points in Figure~\ref{fig_tan_shear}
unlike the cases of Jee et al. (2005a; 2005b), where we excluded a few tangential shear values near the cluster core
to avoid the possible effects of the apparent substructure and blue cluster galaxy contamination. 
In the current cluster, however, both
effects appear to be minor and smaller than the statistical uncertainties.
The SIS fit results in the Einstein radius of the cluster $\theta_E=11.7\arcsec\pm0.4\arcsec$ ($\chi^2/$d.o.f=3.61) 
With $\beta=0.265$ (\textsection\ref{section_source_selection}), the implied velocity dispersion 
is $\sigma_v=(1237\pm22)~\mbox{km}~\mbox{s}^{-1}$. Assuming energy equi-partition between gas and dark matter, we
can also convert this velocity dispersion to the {\it lower} limit of the gas temperature $T_X=(9.4\pm0.3)$ keV.
Although the SIS model is in general a good description of cluster mass profiles at small radii, numerical
simulations show that the mass density of relaxed clusters at large radii should drop faster than $\propto r^{-2}$. 
Also, near the cluster core it is believed that the density profile is less steep than $\propto r^{-2}$.
Therefore,
modified density profiles such as an NFW model are a preferred choice over the traditional SIS in the description
of cluster mass profiles.
For the NFW fit, we obtain $r_s=78\arcsec\pm19\arcsec$ ($\sim604$ kpc) and $c=2.7\pm0.3$ ($\chi^2/$d.o.f=1.11).
The comparison of the reduced $\chi^2$ values and the best-fit results in Figure~\ref{fig_tan_shear}
show that the cluster's reduced tangential shear is better described by this NFW model.
The large reduced $\chi^2$ value of the SIS fitting is mainly due to the fact that the observed tangential shear
does not rise as steeply as the SIS prediction at $r\lesssim50\arcsec$. If we assume a non-singular core instead 
[NIS; i.e., $\kappa=\kappa_0/(r^2+r_c^2)^{1/2}$],
the discrepancy is substantially reduced, and 
we obtain  $r_c=9.5\arcsec\pm1.2\arcsec$ ($\sim74$ kpc) and $\kappa_0=7.4\pm0.7$ with $\chi^2$/d.o.f=0.77.

\begin{figure}
\begin{center}
\includegraphics[width=12cm]{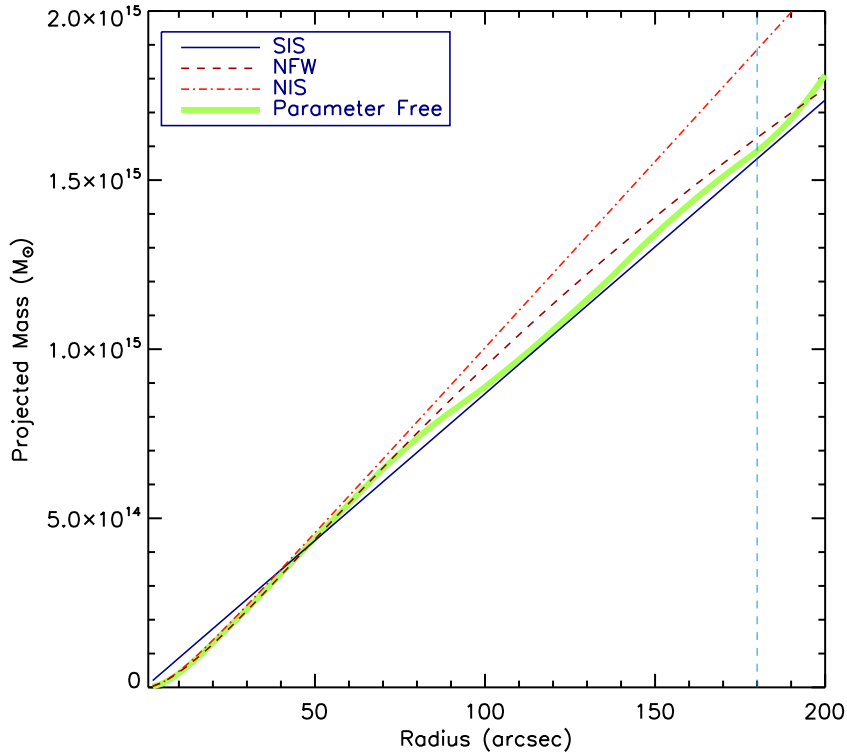}
\caption{Projected mass profile of CL1226. We compare the results from the various methods discussed in \textsection\ref{section_mass_estimation}.
The discrepancy is small over the entire range of the radii shown here except for the NIS model, which, although similar to the other results at 
small radii ($r\lesssim70\arcsec$), gives substantially higher masses at large radii (e.g., $\sim15\%$ higher at $r\sim150\arcsec$). 
Note that for the parameter-free
method the azimuthal average is estimated from a complete circle only at $r\lesssim180\arcsec$ (blue dashed). Error bars are omitted to
avoid clutter. For the SIS result, the mass uncertainties are $\sim6.5$\% (after we
rescale with the reduced $\chi^2$ value) over the entire range. The uncertainty in the NFW mass non-uniformly increases with radii:
approximately 5\%, 10\%, and 15\% of the total mass at $r=50\arcsec, 100\arcsec$, and $200\arcsec$. 
\label{fig_mass_comparison}}
\end{center}
\end{figure}

We compare the projected mass profiles estimated from these results in Figure~\ref{fig_mass_comparison}. Also plotted is
the result based on the two-dimensional mass map of the cluster, for which
we lifted the mass-sheet degeneracy $\kappa \rightarrow \lambda \kappa + 1-\lambda$ by constraining $\bar{\kappa}(150\arcsec<r<200\arcsec)$
to be the same as the value given by the NFW result. 
The discrepancy from difference approaches is small over the entire range of the radii shown here except for 
the NIS model, which, although similar to the other results at 
small radii ($r\lesssim70\arcsec$), gives substantially higher masses at large radii (e.g., $\sim15\%$ higher at $r\sim150\arcsec$). 
Because the $\propto r^{-2}$ behavior of the NIS model at large radii is unrealistic (despite its smallest goodness-of-fit value), 
we do not consider the result as representative of the overall cluster mass profile.
Error bars are omitted to
avoid clutter in Figure~\ref{fig_mass_comparison}. For the SIS result, the mass uncertainties are $\sim6.5$\% (after we
rescale with the reduced $\chi^2$ value) over the entire range. The uncertainty in the NFW mass non-uniformly increases with radii:
approximately 5\%, 10\%, and 15\% of the total mass at $r=50\arcsec, 100\arcsec$, and $200\arcsec$.

\section{DISCUSSION}
\subsection{Comparison with Other Studies}
The X-ray temperature of CL1226 was first measured by Cagnoni et al. (2001) based on short exposure ($\sim10$~ks) $Chandra$ data. 
They obtained an X-ray temperature of $10_{-3}^{+4}$ keV, a $\beta$ index of $0.770\pm0.025$, and a core radius of $r_c=18\arcsec.1\pm0\arcsec.9$.
With an isothermal $\beta$ model assumption, their measurement gives a projected (within a cylindrical volume) mass of $M(r<1~\mbox{Mpc})=(1.4_{-0.4}^{+0.6})\times10^{15} M_{\sun}$, which
is consistent with our result. Joy et al. (2001) used Sunyaev-Zeldovich effect (SZE) observations and determined the mass of the cluster to be 
$M(r<65\arcsec)=(3.9\pm0.5)\times10^{14}~M_{\sun}$
with a SZE temperature of $9.8_{-1.9}^{+4.7}$ keV. This SZE result is also in accordance with our lensing estimate.
With XMM-Newton observations, Maughan et al. (2004) derived a virial mass of $(1.4\pm0.5)\times10^{15}~M_{\sun}$ within $r_{200}=1.66\pm0.34~$Mpc. Our
NFW fitting results gives $r_{200}=1.64\pm0.10~$Mpc and $M(r<r_{200})=(1.38\pm0.20)\times10^{15}~M_{\sun}$, which are in excellent
agreement with the result of Maughan et al. (2004).
Maughan et al. (2007) refined their early study of CL1226 by using both deep XMM-Newton and Chandra observations. From the comprehensive
analysis of the cluster's three-dimensional gas and temperature structure, they obtained $r_{500}=(0.88\pm0.05)~$Mpc and
$M_{500}=5.2_{-0.8}^{+1.0}\times 10^{14}~M_{\sun}$. The new mass is, however, $\sim30$\% lower than
our lensing estimate $M(r<0.88~\mbox{Mpc})=(7.34\pm0.71)\times10^{14}~M_{\sun}$.
This mass discrepancy is interesting because Maughan et al. (2007) noted that their
mass is $\sim30$\% below the $M-T$ relation (Vikhlinin et al. 2006) and also 
the $M-Y_X$ relation (Kravtsov et al. 2006). 
They suggested the possibility that the on-going merger indicated by their temperature map 
may lead to the underestimation of the total mass with X-ray methods.
Our mass reconstruction resolves the cluster core substructure and 
supports the merger scenario. If the merger is indeed responsible for the
underestimation from the X-ray analysis, the apparent improvement in the $M-T$ relation with the lensing estimate
highlights the merits of gravitational lensing for mass estimation, which does not depend on the dynamical state of the system.

\subsection{Stage of the Merger in the Core of CL1226 \label{section_merger}}
Galaxy cluster cores frequently possess merging signatures despite their relaxed morphology on a large scale.
Even the Coma Cluster, long regarded as the archetype of relaxed clusters, has been found to be
composed of many interesting substructures by X-ray and optical observations (e.g., Biviano et al. 1996).
Several lines of evidence strongly suggest that the cluster CL1226 is also undergoing an active merger in the cluster core.
First, the cluster galaxy distribution is bimodal in the core; the dominant number density peak is close to the X-ray center and
the other overdensity is seen to the southwest. Second, the X-ray temperature map
shows that the ICM near this southwestern number density peak is significantly higher ($14\sim18$ keV) than in the neighboring region.
Although this high-temperature region does not stand out in the X-ray surface brightness map, which reveals only a single peak
in spatial agreement with the dominant galaxy number density peak, a scrutiny of both the $Chandra$ and the $XMM-Newton$
images shows that the contours near the X-ray peak is slightly elongated toward the high-temperature region.
Third, our weak-lensing analysis confirms that a substantial mass is associated with this southwestern galaxy number density peak.
Therefore, it is plausible that the hydrodynamic interaction between the two substructures 
is responsible for the high temperature ($14\sim18$ keV) feature. 

It is puzzling that the southwestern weak-lensing mass clump is not detected in the X-ray surface brightness.
As mentioned in \textsection\ref{section_mass_map}, the mass of this substructure is comparable to that of the western
weak-lensing mass peak of MS1054, which is the most massive of the three dominant peaks of the cluster and has
its own distinct X-ray peak.
As a possible cause, we suggest the possibility that the substructure has passed through the other more massive structure
from the eastern side.
If it is a slow encounter, the gas of the system could be severely stripped during this penetration.
A shock, on the other hand, can propagate ahead of the gas core and leave
trails of hot temperature as observed in the current case. 
Of course, the collisionless galaxies and dark matter of the system are expected to survive the core passthrough as observed unlike
the gas system. A famous example is the ``bullet'' cluster 1E0657-56 at $z\simeq0.3$ (Clowe et al. 2006; Bradac et al. 2006).

We noticed a very similar case
in our weak-lensing and X-ray study of MS1054 (Jee et al. 2005b). The eastern substructure of 
MS1054, whose presence is clear both in the cluster galaxy and dark matter distribution,
is conspicuously absent in the X-ray observations. Moreover, the MS1054 temperature map of Jee et al. (2005b) shows that
the gas in the region of the eastern halo is notably higher ($\gtrsim10$ keV) than the average temperature ($\sim8.9$ keV).
Hence in Jee et al. (2005b) we proposed that the cluster galaxies now observed in the eastern mass clumps might have
passed through the central mass clump from the southwest. Intriguingly, a significant fraction of the star forming galaxies
(four out of the five brightest IR galaxies) are found to exist near the eastern mass clump of MS1054 (Bai et al. 2007). 
The result can be interpreted as indicating a recent star formation triggered by this hypothesized merger.
Although we have not performed a parallel study for the star-formation properties of the CL1226 galaxies yet, the existing
features in the X-ray surface brightness, gas temperature, mass, and galaxy distribution is suggestive of 
the similar merger scenario.

Is the above post-merger picture the unique scenario that explains the observed features?
If the peculiar temperature structure of CL1226 were absent or we can attribute it to something else,
one can also consider the possibility that the secondary mass clump might not be massive enough to produce X-ray emission, but
still detected in weak-lensing simply because it forms a line-of-sight superposition with the already massive dark matter 
halo of the primary cluster.
Obviously, in this hypothesized configuration the background density can boost the lensing signal even if 
the mass of the southwestern clump by itself is not significant. To explore this possibility quantitatively, we
estimated the expected X-ray temperature of the secondary clump for this scenario in the following way.
First, we created
a new radial density profile of the main cluster from the mass map by excluding the azimuthal range that contains the southewestern substructure.
Second, we subtracted this new radial density profile from the original mass map.
Finally, we measured
the mass of the secondary clump from this subtracted mass map. Of course, this procedure 
overestimates the contribution (i.e., boosting effect) of the primary cluster because, even if we avoided the southwestern region in the
creation of the new radial profile, the $\kappa$ value in the other azimuthal range is still the sum of the two subclusters.
Hence, here we assume an extreme case, where the primary cluster is dominant in mass.
Then, within $r=30\arcsec$ ($\sim232$ kpc), the total projected mass of the secondary cluster would be $\sim5\times10^{13} M_{\sun}$.
Assuming isothermality with $\beta_X=0.7$, this mass is translated into $T_X \sim 2.6$ keV.
Given the depth ($\sim72$ ks for each detector) of the observation, a subcluster with this cool core
would have been easily identified in the $XMM-Newton$ image of CL1226 (Maughan et al. 2007).
Even for the relatively shallow  $Chandra$ data ($\sim50$ ks when the two datasets ObsID 3180 and 5014 are combined), 
we predict $\sim200$ counts within a $r=10\arcsec$ aperture, which would give a significance of $\sim8~\sigma$.
Therefore, even if we disregard the temperature structure, we are not likely to be observing a very low mass
cluster whose lensing efficiency is enhanced due to the primary cluster halo.

\section{SUMMARY AND CONCLUSIONS}

We have presented a weak-lensing study of the galaxy cluster CL1226 at $z=0.89$. The cluster
is a very interesting, rare system because, despite its high redshift, it has a relaxed morphology in X-ray surface brightness 
and an unusually high gas temperature. Our HST/ACS-based weak-lensing analysis of the cluster provides 
the dark matter distribution in unprecedented detail and allows us to measure the mass profiles out to the virial radius
of the cluster.

Our two-dimensional mass reconstruction shows that on large scales the dark matter distribution
is consistent with a relaxed system with no significant substructures. However, viewed in detail, the cluster core
is resolved into two mass clumps.
This bimodality of the core mass structure is also seen in the cluster galaxy distribution.
The dominant mass clump lies close to the BCG whereas the other less massive one is located $\sim40\arcsec$ to the southwest.
This secondary mass clump does not stand out in the X-ray surface brightness although the temperature
of the region is much higher than in the rest of the cluster. When the significant mass associated with
the substructure is considered, the absence of the corresponding X-ray excess in the region is puzzling.
Therefore, we propose that
we may be observing the system after the less massive subcluster passed through the main cluster.
It is possible that the X-ray gas of the less massive system might have been stripped due to the ram pressure.
The slight elongation of the X-ray peak toward the southwestern mass clump is also supportive of this scenario.
These features are similar to
the ones that we observed in MS1054, another massive galaxy cluster at $z=0.83$, where we proposed a 
similar possibility.

We measure significant shear signals out to the field boundary ($\sim200\arcsec$), which indicates
that the cluster is indeed massive as already implied by its high X-ray temperature.
Fitting an NFW profile to the reduced tangential shears gives $r_{200}=1.64\pm0.10~$Mpc and $M(r<r_{200})=(1.38\pm0.20)\times10^{15}~M_{\sun}$, where
the error bars include both statistical and cosmic-shear induced systematic uncertainties.
Although the predicted velocity dispersion and X-ray temperature from the lensing result are
consistent with previous work, our cluster mass is $\sim30$\% higher than the recent XMM-$Newton$ and $CHANDRA$
analysis of Maughan et al. (2007), who interestingly pointed out that their mass estimate is
$\sim30$\% below the $M-T$ and $Y_X-T$ scaling relations.
If the on-going merger is indeed the cause of the underestimation of the total mass with the X-ray method (Maughan et al. 2007),
the apparent improvement in the $M-T$ relation with the lensing estimate
highlights the advantages of gravitational lensing for mass estimation, which does not depend on the dynamical state of the system.

Our weak-lensing study confirms that CL1226 is indeed the most massive cluster at $z>0.6$ known to date. In the
hierarchical structure formation paradigm, a CL1226-like cluster is extremely rare at $z=0.89$.
Because the abundance of such a massive system is sensitive to the matter density ($\Omega_M$) and its 
fluctuation ($\sigma_8$), the current result and future lensing measurements of other high-redshift
massive clusters [e.g., XMMU J2235.3-2557 at $z=1.39$ (Mullis et al. 2005) and XCS J2215.9-1738 at $z=1.45$ 
(Stanford et al. 2006; Hilton et al. 2007) ] will provide useful constraints on the normalization of the power spectrum.

M. James Jee acknowledges support for the current research from the TABASGO foundation presented in the form of
the Large Synoptic Survey Telescope Cosmology Fellowship. 
We thank Ben Maughan for allowing us to use his X-ray temperature map of CL1226.

\begin{figure}
\plotone{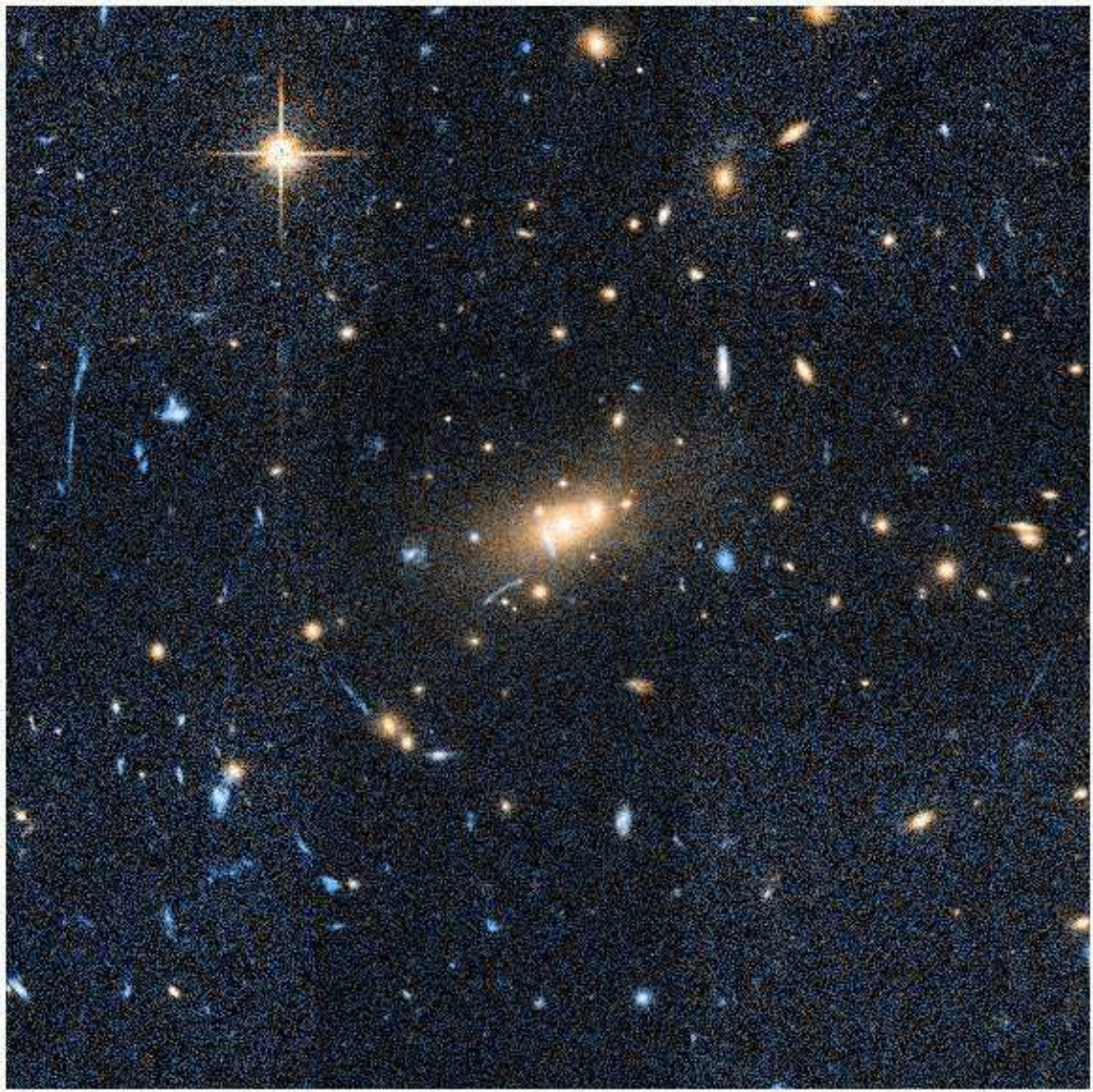}
\caption{The central $30\arcsec\times30\arcsec$
region, which is marked with a yellow square in Figure~\ref{fig_cl1226}, is blown up to show details. The cluster red-sequence can be easily identified 
by their distinct colors (yellow in this rendition). Also, many strong-lensing features (i.e., blue arcs) are observed.
\label{fig_cl1226_center}}
\end{figure}

\end{document}